\documentclass[fleqn,twoside]{article}
\usepackage{espcrc2,epsf,latexsym}

\newcommand{\msbar}{\mbox{\tiny $\overline{MS}$}}        
\newcommand{\plaq}{\Box}                                 

\def\lsim{\mathrel{\rlap{\lower4pt\hbox{\hskip1pt$\sim$}}
    \raise1pt\hbox{$<$}}}                
\def\gsim{\mathrel{\rlap{\lower4pt\hbox{\hskip1pt$\sim$}}
    \raise1pt\hbox{$>$}}}                

\title{
       \vspace{-3.65cm}                                     %
       {\normalsize DESY 04-167}      \\[-0.2cm]            
       {\normalsize Edinburgh 2004/17}\\[-0.2cm]            
       {\normalsize LTH 631}          \\[-0.2cm]            
       {\normalsize LU-ITP 2004/030}\\[-0.2cm]              
       {\normalsize September 2004}  \\                     
       \vspace{1.32cm}                                      
       Determination of Lambda in quenched and full QCD -- an update%
            \thanks{Talk given by R. Horsley at Lat04,
                    Fermilab, USA.}}                       

\author{M. G\"ockeler%
           \address{Institut f\"ur Theoretische Physik, Universit\"at
                    Leipzig, D-04109 Leipzig, Germany}
                       \hspace{-0.25cm} $^,$\hspace{-0.15cm}
           \address{Institut f\"ur Theoretische Physik, Universit\"at
                    Regensburg, D-93040 Regensburg, Germany},
        R. Horsley%
           \address{School of Physics,
                    University of Edinburgh, Edinburgh EH9 3JZ, UK},
        A.~C. Irving%
           \address{Department of Mathematical Sciences,
                    University of Liverpool, Liverpool L69 3BX, UK},
        D. Pleiter%
           \address{John von Neumann Institute NIC / DESY Zeuthen,
                    D-15738 Zeuthen, Germany},
        P.~E.~L. Rakow%
           $^{\rm d}$,
        G. Schierholz%
           $^{\rm e,}$%
           \address{Deutsches Elektronen-Synchrotron DESY,
                    D-22603 Hamburg, Germany}
        and
        H. St\"uben%
           \address{Konrad-Zuse-Zentrum f\"ur Informationstechnik Berlin,
                    D-14195 Berlin, Germany}
        \\ -- {\it QCDSF--UKQCD} Collaboration }

\begin{document}

\begin{abstract}
We present an update on our previous determination of the Lambda parameter
in QCD. The main emphasis is on results for two flavours
of light dynamical quarks, where we can now almost double the amount
of data used, including values at smaller lattice spacings.
The calculations are performed using $O(a)$ improved Wilson fermions.
Little change is found to previous numerical values.
\end{abstract}

\maketitle

\setcounter{footnote}{0}




The $\Lambda$ parameter is one of the fundamantal parameters of QCD,
setting the scale for the running coupling constant $\alpha_s$. In this
contribution we shall update our previous work, \cite{booth01a},
both for quenched ($n_f=0$) and unquenched ($n_f=2$) $O(a)$ improved
Wilson (`clover') fermions. Specifically we are now able to use for
\begin{itemize}
   \item quenched fermions, the force scale $r_0/a$ up to $\beta=6.92$,
         \cite{necco01a} (previously $\beta \le 6.4$), 
   \item unquenched fermions, improved statistics and additional
         quark masses at the previous $\beta$ values of
         $5.20$, $5.25$, $5.29$ for $r_0/a$ together with new results at 
         $\beta=5.40$ (at three quark masses).
\end{itemize}

\begin{figure}[htb]
   \epsfxsize=7.00cm \epsfbox{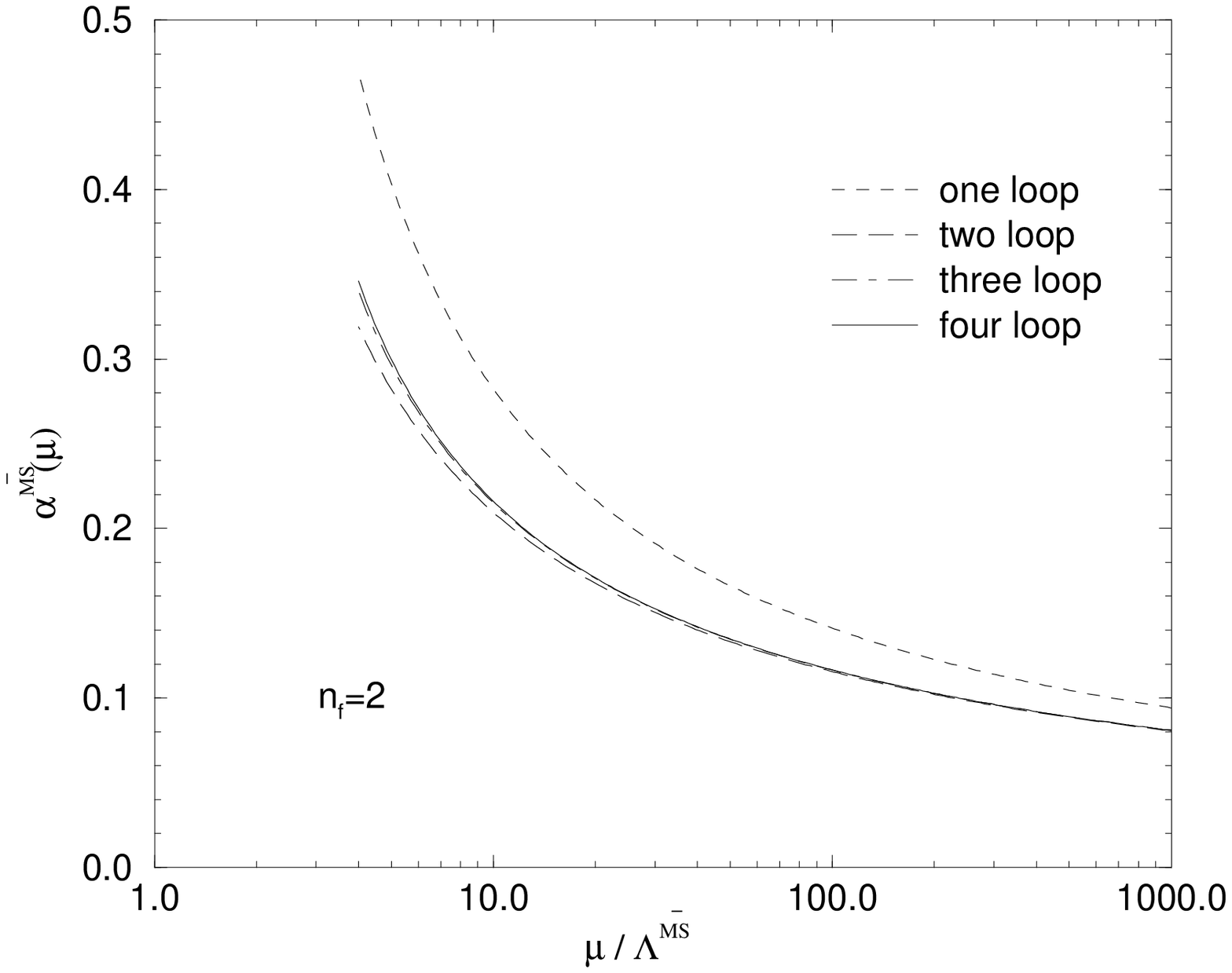}
   \vspace*{-0.30in}
   \caption{\footnotesize{\it $\alpha_s^{\msbar}(\mu)$ versus
            $\mu/\Lambda^{\msbar}$ for $n_f=2$.}}
   \vspace*{-0.20in}
   \label{fig_alpha_nf2_mu}
\end{figure}

The `running' of the QCD coupling constant as the scale changes is
controlled by the $\beta$-function
\begin{eqnarray}
   {\partial g_{\cal S}(M) \over \partial \log M }
      = \beta^{\cal S} \left(g_{\cal S}(M)\right) \,,
                                          \nonumber
\end{eqnarray}
where, perturbatively
\begin{eqnarray}
   \beta^{\cal S} \left(g_{\cal S}\right)
       = - b_0g_{\cal S}^3 - b_1g_{\cal S}^5
         - b_2^{\cal S}g_{\cal S}^7 
         - b_3^{\cal S}g_{\cal S}^9 - \ldots \,,
                                          \nonumber
\end{eqnarray}
renormalisation having introduced a scale $M$ together with a scheme
$\cal S$. Integrating this equation gives
\begin{eqnarray}
   {\Lambda^{\cal S} \over M}
     &=& \exp{\left[ - {1\over 2b_0 g_{\cal S}(M)^2}\right]} 
         \left[b_0 g_{\cal S}(M)^2 \right]^{- {b_1\over 2b_0^2}}
         \times
                                          \nonumber     \\
     & & \hspace*{-0.50in}\quad \exp{\left\{ - \int_0^{g_{\cal S}(M)} d\xi
         \left[ {1 \over \beta^{\cal S}(\xi)} +
                {1\over b_0 \xi^3} - {b_1\over b_0^2\xi} \right]\right\} } \,,
                                          \nonumber     \\
     &\equiv& F^{\cal S}(g_{\cal S}(M)) \,,
                                          \nonumber  
\end{eqnarray}
where $\Lambda^{\cal S}$, the integration constant, is the fundamental
scheme dependent QCD parameter. Results are usually given in the
$\overline{MS}$ scheme, with the scale $M$ being denoted by $\mu$.
In this scheme the first four $\beta$-function coefficients are known,
$b_3^{\msbar}$ being found in \cite{vanritbergen97a}.
The running coupling  $\alpha^{\msbar}_s(\mu) \equiv g_{\msbar}(\mu)^2/4\pi$
is plotted in Fig.~\ref{fig_alpha_nf2_mu} for $n_f=2$, by solving
the previous equation (numerically) using successively more and more
coefficients of the $\beta$-function. The figure shows an apparently
fast convergent series (cf 3- to 4-loop), certainly in the range we
are interested in, $\mu/\Lambda^{\msbar} \sim 8$. A very similar result
holds for $n_f=0$ but with slightly lower curves.




On the lattice we also have a $\Lambda$ parameter,
\begin{eqnarray}
   a \Lambda^{\plaq}  = F^{\plaq}(g_{\plaq}(a)) \,,
                                          \nonumber  
\end{eqnarray}
where to help convergence of lattice perturbative expansions we use
$g_{\plaq}^2 \equiv g^2(a) / u_0^4$ with $u_0^4$ the average
plaquette value. To calculate $\Lambda^{\msbar}$,
we shall compute $g_{\msbar}$ at some appropriate scale $\mu^*$ from 
$g_{\plaq}(a)$ and then using the $r_0$ scale, extrapolate
\begin{eqnarray}
   r_0 \Lambda^{\msbar} 
      \equiv  \left( r_0 \over a \right) F^{\msbar}( g_{\msbar}(\mu^*)) 
              a \mu^* \,,
                                          \nonumber  
\end{eqnarray}
to the continuum limit.

Equating lattice and continuum expressions
\begin{eqnarray}
   [F^{\msbar}(g_{\msbar}(\mu))]^{-1} 
       = a\mu { \Lambda^{\plaq} \over \Lambda^{\msbar} }
             [F^{\plaq}(g_{\plaq}(a))]^{-1} \,,
                                          \nonumber  
\end{eqnarray}
and expanding as
\begin{eqnarray}
   {1 \over g_{\msbar}^2(\mu)}
     &=& {1 \over g_{\plaq}^2(a)} +
                                          \nonumber     \\
     & &  \hspace*{-0.50in} [2b_0 \ln a\mu - t^{\plaq}_1]
          + [2b_1 \ln a\mu - t^{\plaq}_2] g_{\plaq}^2(a)
          \ldots \,,
                                          \nonumber  
\end{eqnarray}
gives $t_1^{\plaq} = 2 b_0 \ln \Lambda^{\msbar}/ \Lambda^{\plaq}$
and $b_2^{\plaq} = b_2^{\msbar} + b_1t^{\plaq}_1 - b_0t^{\plaq}_2$.
For (hopefully) good convergence of this series we choose the scale so
that the $O(1)$ term vanishes, $a\mu^* = \exp ( t_1^{\plaq} / 2b_0 )$.

For $t_1^{\plaq}$ the general expression is known for $n_f$, $c_{sw}$
and linear terms in $n_fam_q$, while for $t_2^{\plaq}$ the $n_fam_q$
dependence is not known, \cite{booth01a} and references therein.
We can estimate the scales as $\mu^* = 2.63/a$, $n_f=0$ and $\mu^* \sim 1.4/a$
for $n_f=2$. $t_3^{\plaq}$ (the $g_{\msbar}(\mu)^4$, $\ln a\mu$ independent
term) is not known. So equivalently $b_3^{\plaq}$ is not known. However a
Pad{\'e} estimate gives $b_3^{\cal S} \approx (b_2^{\cal S})^2/b_1$,
and is small and in reasonable agreement with the known coefficient
in the $\overline{MS}$ scheme, \cite{booth01a}. Assuming this also
holds for $b_3^{\plaq}$ gives little change to the results presented here.
For complete $O(a)$ cancellation, \cite{luscher96a}, we need
$\tilde{g}^2 = g^2 ( 1 + b_g am_q )$ where perturbatively
$b_g = 0.01200n_f g^2 + O(g^4)$, which with
$c_{sw} = 1 + O(g^2)$ then gives no mass dependence in $t_1^{\plaq}$.
This indicates little quark mass dependence in the fit formulae (indeed
there is more in the numerical data).
Finally to further improve the convergence
of the series, we tadpole improve the $t_i^{\plaq}$ coefficients
$c_{sw}^{TI} = c_{sw} u_0^3$ (for $t_1^{\plaq} + t_2^{\plaq}g_{\plaq}^2$)
further reducing the size of the $n_f$ term in $t_2^{\plaq}$.

\begin{figure}[htb]
   \vspace*{-0.10in}
   \epsfxsize=7.00cm \epsfbox{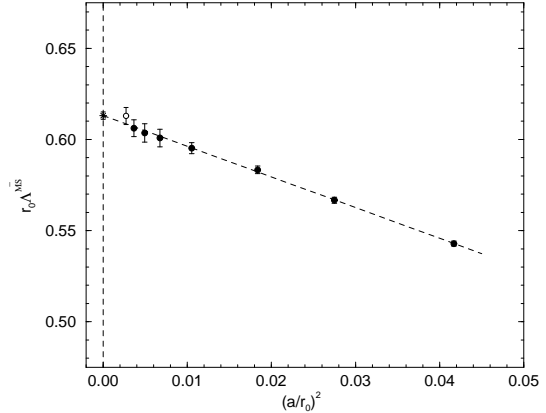}
   \vspace*{-0.30in}
   \caption{\footnotesize{\it $r_0\Lambda^{\msbar}$ versus $(a/r_0)^2$
            for $n_f=0$, together with a linear extrapolation to $a=0$.
            The last point has not been included in the fit.}}
   \vspace*{-0.30in}
   \label{fig_r0lamMSbar_b0+b1+b2+b3_nf0}
\end{figure}

In Fig.~\ref{fig_r0lamMSbar_b0+b1+b2+b3_nf0} we show the quenched ($n_f=0$)
results.
The data lies on a straight line (as a function of $(a/r_0)^2$)
at least over $a^{-1} \sim 2$ -- $6.5\,\mbox{GeV}$  or
$\mu \sim 5$ -- $17 \, \mbox{GeV}$, using the value for $r_0$ of
$r_0 = 0.5\, \mbox{fm}$. This gives a result of
$r_0\Lambda^{\msbar} = 0.613(2)(25)$ or
$\Lambda^{\msbar}(0) = 242(1)(10)\,\mbox{MeV}$
where the first error is statistical and to estimate the systematic
uncertainty, the second error takes a
$g^4 \, \mbox{coeff.} = 25\% \times g^2 \, \mbox{coeff.}$ (which is
very much greater than when using the Pad{\'e}
$b^{\plaq}_3$ estimate).

\begin{figure}[htb]
   \vspace*{-0.30in}
   \epsfxsize=7.00cm \epsfbox{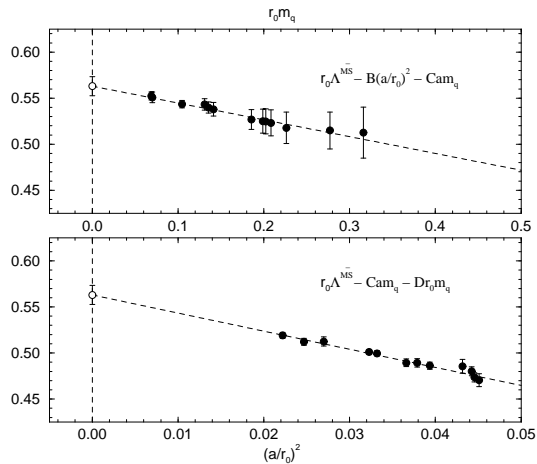}
   \vspace*{-0.30in}
   \caption{\footnotesize{\it $r_0\Lambda^{\msbar}$ versus $r_0m_q$
            (upper picture) and versus $(a/r_0)^2$ (lower picture)
            for $n_f=2$, together with appropriate extrapolations
            ($am_q$ from \cite{gockeler04a}).}}
   \vspace*{-0.30in}
   \label{fig_r0lamMSbar_nf2_040614_1529_lat04}
\end{figure}

For unquenched ($n_f=2$) fermions, due to the sea quark, the fit ansatz
is not so simple as we must consider both chiral and continuum
extrapolations. We take for finite $a$,
$a\Lambda^{\msbar}|_{m_q \not= 0, a \not= 0}
            = a\Lambda^{\msbar}|_{m_q = 0, a \not= 0} + Dam_q + \ldots$
or $r_0\Lambda^{\msbar}|_{m_q \not= 0, a \not= 0}
            = r_0\Lambda^{\msbar}|_{m_q = 0, a \not= 0} + Dr_0m_q +\ldots$.
After chiral extrapolation we would thus expect
$r_0\Lambda^{\msbar}|_{m_q = 0, a \not= 0}
            = r_0\Lambda^{\msbar}|_{m_q = 0, a = 0} + B (a/\rho)^2 + \ldots$
with $\rho \equiv r_0|_{m_q=0}$. Together with
$(a/r_0)^2 = (a/\rho)^2 + Eam_q + \ldots$ this gives our fit ansatz as
\begin{eqnarray}
   r_0\Lambda^{\msbar} = A + B (a/r_0)^2 + Cam_q + Dr_0m_q \,.
                                          \nonumber  
\end{eqnarray}
So by subtracting out the $B$ and $C$ terms from $r_0\Lambda^{\msbar}$
we can consider the chiral extrapolation and similarly by subtracting out
the $C$ and $D$ terms we may consider the continuum extrapolation%
\footnote{An alternative procedure is first to extrapolate both $r_0/a$
and $u_0^4$ to the chiral limit, evalute $r_0\Lambda^{\msbar}$
and then extrapolate to the continuum limit; this gives similar results,
\cite{gockeler04a}.}.
In Fig.~\ref{fig_r0lamMSbar_nf2_040614_1529_lat04}
we show the results. $a^{-1}$ ranges at least over
$a^{-1} \sim 2$ -- $3\,\mbox{GeV}$ or $\mu \sim 3$ -- $4 \, \mbox{GeV}$
This gives a result of $r_0\Lambda^{\msbar} = 0.563(10)(70)$ or
$\Lambda^{\msbar}(2) = 222(4)(28)\,\mbox{MeV}$
where again the first error is statistical and the second error is
obtained by taking a
$g^4 \, \mbox{coeff.} = 25\% \times g^2 \, \mbox{coeff.}$
which again is much larger than the error found when using a Pad{\'e}
$b^{\plaq}_3$ estimate, setting $c_{sw} = 1 + O(g^2)$ or including an
additional $(am_q)^2$ fit term. Note that this result is consistent
with that obtained in \cite{bali02a}.

\begin{figure}[htb]
   \epsfxsize=7.00cm \epsfbox{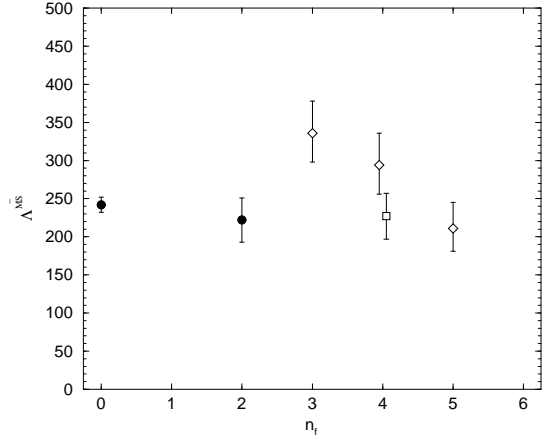}
   \vspace*{-0.30in}
   \caption{\footnotesize{\it $\Lambda^{\msbar}(n_f)$ versus $n_f$
            The open diamonds are from \cite{bethke02a},
            using $\alpha_s^{\msbar}(M_Z) = 0.1183(27)$ ($n_f=5$) to
            match to $n_f=4$ and $n_f=3$, while the open square is
            from \cite{blumlein04a}. The filled circles
            are the results reported here.}}
   \vspace*{-0.30in}
   \label{fig_lambda_nf_bethke+blum}
\end{figure}

Finally in Fig.~\ref{fig_lambda_nf_bethke+blum} we present results for 
different $n_f$. Our result lies somewhat low in comparison with
phenomenological results. Alternatively using the matching procedure
as in \cite{booth01a} we find for
$n_f=5$, $\alpha^{\msbar}_s(m_Z) = 0.1084(6)(38)$.



\section*{ACKNOWLEDGEMENTS}

The numerical calculations have been performed on the Hitachi SR8000 at
LRZ (Munich), on the Cray T3E at EPCC (Edinburgh) under
PPARC grant PPA/G/S/1998/00777, \cite{allton01a},
on the Cray T3E at NIC (J\"ulich) and ZIB (Berlin),
as well as on the APE1000 and Quadrics at DESY (Zeuthen).
This work is supported in part by
the EU Integrated Infrastructure Initiative Hadron Physics (I3HP) 
and by the DFG (Forschergruppe Gitter-Hadronen-Ph\"anomenologie).



\end{document}